O.V.Bespalova, E.A.Romanovsky, T.I.Spasskaya

# NUCLEON-NUCLEUS REAL POTENTIAL OF WOODS-SAXON SHAPE BETWEEN –60 AND +60 MEV FOR THE $40 \leq A \leq 208$ NUCLEI

M.V.LOMONOSOV MOSCOW STATE UNIVERSITY
D.V.SCOBELTSYN INSTITUTE OF NUCLEAR PHYSICS

O.V.Bespalova, E.A.Romanovsky, T.I.Spasskaya

# NUCLEON-NUCLEUS REAL POTENTIAL OF WOODS-SAXON SHAPE BETWEEN –60 AND +60 MEV FOR THE $40 \leq A \leq 208$ NUCLEI

**O.V.Bespalova, E.A.Romanovsky, T.I.Spasskaya**

e-mail: besp@hep.sinp.msu.ru

# NUCLEON-NUCLEUS REAL POTENTIAL OF WOODS-SAXON SHAPE BETWEEN –60 AND +60 MEV FOR THE $40 \le \mathrm{A} \le 208$ NUCLEI

**Abstract**

The nucleon differential elastic scattering cross sections, the total proton reaction cross sections, and the single-particle energies of nucleon bound states for $^{40}$Ca, $^{90}$Zr, and $^{208}$Pb nuclei are reanalyzed in terms of the dispersive optical model at energies ranging from –75 MeV to 60 MeV. The resultant real effective Woods-Saxon potential, which corresponds to the dispersive potential, is studied as dependent on A, Z, and E and on projectile specie (proton or neutron). For the first time, a parameterization of the Woods-Saxon real part of the nucleon-nucleus optical potential is proposed for the $40 \le A \le 208$ nuclei at energy ranging from –60 MeV to +60 MeV, including a range near the Fermi energy. The parameterization reflects the dispersion relation between the real and imaginary parts of the optical model potential through the energy dependence of the radius parameter of the real part of the potential. The method to determine the imaginary part of the optical model potential, which is symmetrical relative to the Fermi energy, is also proposed for the $40 \le A \le 208$ nuclei. The differential elastic scattering cross sections, the total neutron interaction cross sections, the total proton reaction cross sections, and the single-particle energies of the nucleon bound states calculated in terms of the proposed nucleon-nucleus potential parameterization for some of the n,p+A ($40 \le A \le 208$) systems are compared with the available experimental data, yielding a fairly good agreement.

## 1.Introduction

Nuclear optical-model (OM) potential is a powerful tool to analyze elastic scattering of pions, nucleons, and heavier particles. OM potential is widely used to generate distorted waves with the view of analyzing nuclear reactions, so the global OM potentials are very important in the numerous cases where the distorted waves have to be used for the energies and nuclei that are not supported by any available experimental scattering data. The global OM potentials can also be used to predict the scattering data for unstable nuclei, for which no experimental data are available, and to test microscopically calculated potentials.

The last two decades have seen great advances in getting a proper formulation of the nuclear mean field unified for positive and negative energies (Mahaux and Sartor, 1991a). The unified description of the nuclear mean field is based on the concept that the shell-model potential is complex and is the continuation of the OM potential. The extrapolation of the OM potential from positive to negative energies is based on the dispersion relations between the real and imaginary parts of the OM potential. The dispersion relations follow from the causality requirement to the effect that an outgoing wave cannot be emitted before arriving an incident wave. The unified mean field potential is determined by the dispersive OM analysis of the scattering data and the single-particle energies of bound states. The central part of the local dispersive OM potential is a sum of its "static" and "dynamic" components. The "static" component, which depends smoothly on energy due to the local approximation, is the so-called Hartree-Fock (HF) potential $V_{HF.}$ The “dynamic” component depends strongly on energy in the energy range near the Fermi energy $E_F$ and is supposed to carry information about the correlation (at $E<E_F$) and dynamic polarization (at $E>E_F$) effects. Dispersion relations hold between the real and imaginary parts, $\Delta V$ and $W$, of the dynamic component because of the supposed analytical properties of the nuclear potential.

Having been applied to the various given n,p+nucleus systems, dispersive OM analysis yielded an agreement between the calculated and experimental scattering data (differential elastic scattering cross sections, total proton reaction cross sections, total neutron interaction cross sections, polarization) and bound state data (single-particle energies, spreading widths, root-mean-square radii, occupation probabilities, spectroscopic factors, spectral functions).

This study presents the nucleon global OM potential with the Woods-Saxon real part, which reflects the dispersion relation, for the $40 \le A \le 208$ nuclei at nucleon energies from –60 to +60 MeV, including the near-Fermi energy range. Section 2 describes the dispersive OM analysis as applied to the scattering and bound-state data of the n,p+$^{40}$Ca, $^{90}$Zr, and $^{208}$Pb systems. In Section 3, the Woods-Saxon real part of the global OM potential derived from dispersive OM potential is presented for the $40 \le A \le 208$ nuclei at energies from –60 to + 60 MeV. Besides, a method for calculating the parameters of the imaginary part, which is symmetric relative to the Fermi energy, is proposed. Section 4 compares the scattering and bound-state data calculated using proposed global OM potential with the available experimental and theoretical results. Finally, Section 5 presents a discussion and brief description of the main features of the proposed global OM potential.

## 2. Dispersive optical-model analysis

The nucleon-nucleus potential of the conventional (nondispersive) and dispersive OMs is

$$U(r,E) = V_C(r) - U_p(r,E) - U_{so}(r,E)\,, \tag{1}$$

where $V_C(r)$ is the Coulomb potential, which is taken to be that of a uniform charged sphere of radius $R_C = r_C \cdot A^{1/3}$; $U_p(r,E)$ is the central potential; $U_{so}(r,E)$ is the spin-orbit potential, whose radial dependence is that of the Thomas formfactor. The central part of the conventional OM potential is defined to be

$$U_p(r,E) = V(E)\,f(r,r_V,a_V) + iW_s(E)\,f(r,r_s,a_s) - i4a_d W_d(E)\frac{d}{dr}f(r,r_d,a_d)\,, \tag{2}$$

where $f(r,r_i,a_i) = \dfrac{1}{1+\exp\left[(r - r_i \cdot A^{1/3})/a_i\right]}$ ($i$=$V$, $s$, $d$) is the Woods-Saxon function; the subscripts $s$ and $d$ stand for the volume and surface parts of the imaginary potential, respectively. In the dispersive OM analysis, the central real potential is represented by the sum of three terms, namely, the Hartree-Fock potential $V_{HF}$ and the volume, $\Delta V_s$ , and surface, $\Delta V_d$ , dispersive components:

$$\begin{aligned} U_p(r,E) = {} & V_{HF}(E)\,f(r,r_{HF},a_{HF}) + \Delta V_s(E)\,f(r,r_s,a_s) - 4a_d \Delta V_d(E)\frac{d}{dr}f(r,r_d,a_d) + \\ & + iW_s(E)\,f(r,r_s,a_s) - i4a_d W_d(E)\frac{d}{dr}f(r,r_d,a_d) \end{aligned}\,. \tag{3}$$

The dispersive components can be determined from a dispersion relation if the appropriate components of the imaginary part of the OM potential are known throughout the energy range:

$$\Delta V_{s(d)}(r,E) = (E_F - E)\frac{P}{\pi}\int_{-\infty}^{\infty}\frac{W_{s(d)}(r,E')}{(E'-E_F)\cdot(E-E')}dE' \tag{4}$$

(P stands for the principal value). The geometrical shapes of the imaginary parts (volume and surface) and the corresponding dispersive terms are the same if the imaginary potential parameters $r_s$, $r_d$, $a_s$, $a_d$ are independent of energy. The details of the dispersive OM analysis can be found in the review (Mahaux and Sartor, 1991a).

The OM potential can conveniently be characterized by the volume integrals per nucleon:

$$J_i(E) = \frac{4\pi}{A}\int_0^{\infty} V_i(r,E) r^2 dr, \qquad i=R, HF,$$

$$J_{\Delta V_i}(E) = \frac{4\pi}{A}\int_0^{\infty} \Delta V_i(r,E) r^2 dr, \; J_i(E) = \frac{4\pi}{A}\int_0^{\infty} W_i(r,E) r^2 dr \qquad i = s, d, \tag{5}$$

and

$$J_I(E) = J_s(E) + J_d(E),$$

where $I$ stands for total (volume plus surface) imaginary potential.

The dispersion relation (4) is also valid for the volume integrals per nucleon of the respective parts of the OM potential:

$$J_{\Delta V_i}(E) = (E_F - E)\frac{P}{\pi}\int_{-\infty}^{\infty} \frac{J_i(E')}{(E' - E_F)\cdot(E - E')} dE', \qquad i = s, d, I. \tag{6}$$

In this work, the Jeukenne-Mahaux (JM) formula (Jeukenne and Mahaux, 1983) is used to express the energy dependence of the volume integrals per nucleon of the imaginary potential:

$$J_i^{JM}(E) = \alpha \frac{(E - E_F)^4}{(E - E_F)^4 + \beta_i^{\,4}}, \text{ where } i = I, s, \tag{7}$$

$$J_d^{JM}(E) = J_I^{JM}(E) - J_s^{JM}(E).$$

In this case, the dispersive components can be calculated analytically:

$$J_{\Delta V}(E) = \frac{\alpha\beta_I (E - E_F)[(E - E_F)^2 + \beta_I^{\,2}]}{\sqrt{2}[(E - E_F)^4 + \beta_I^{\,4}]},$$

$$J_{\Delta V_s}(E) = \frac{\alpha\beta_s (E - E_F)[(E - E_F)^2 + \beta_s^{\,2}]}{\sqrt{2}[(E - E_F)^4 + \beta_s^{\,4}]} \tag{8}$$

$$J_{\Delta V_d}(E) = J_{\Delta V}(E) - J_{\Delta V_s}(E).$$

In order to determine the global parameterisation of the nucleon-nucleus OM potential at energies from –60 to +60 MeV, including the near-Fermi energy range, we reanalysed the scattering and bound state data for the n,p+$^{40}$Ca,$^{90}$Zr,$^{208}$Pb systems in terms of the dispersive OM. These nuclei have been regarded as the nuclei of choice to test the dispersive OM analysis technique (Mahaux and Sartor, 1991a, 1991b, 1994; Wang et al., 1993; Delaroche et. al., 1989; Chiba et al., 1992). The present analysis is based on the progress in specifying the global potentials of the conventional OM, in particular on the most reliable global parameterization of CH89 (Varner et al. 1991). The latter is quite appropriate when applied to the 40<A<209 target-nucleus mass range and the laboratory 10-65 MeV nucleon energy range. The CH89 parameterization was used to determine the parameter $\alpha$ in Eq. (7). The only difference is that the diffuseness of the volume and surface parts of the imaginary potential $a_s^{CH89} = a_d^{CH89} = 0.69$ fm was taken from (Romanovsky et al., 1998). In the latter paper, the total proton reaction cross sections $\sigma_r$ measured to within a 3% accuracy (Carlson 1996) for the stable nuclei $^{40}$Ar, $^{40,42,44,48}$Ca, $^{50,52,53,54}$Cr, $^{51}$V, $^{54,56,57}$Fe, $^{59}$Co, $^{58,60,62,64}$Ni, $^{63,65}$Cu, $^{64,66,68}$Zn, $^{89}$Y, $^{90,92,94,96}$Zr, $^{98,110}$Mo, $^{108,110,112,114,116}$Cd, $^{112,114,116,118,120,122,124}$Sn, $^{140}$Ce, and $^{208}$Pb were analyzed in terms of CH89. The CH89 parameters were fixed, except for the diffuseness parameter $a_s=a_d$, which was varied to fit $\sigma_r$ (Carlson, 1996). The resultant diffuseness $a_s^*=a_d^*$ was found to be below the 0.69 fm value implied by CH89 (the nuclei-averaged $a_s^*=a_d^*=0.63$ fm) and to correlate with the shell structure of a specific nucleus. In the present paper, the diffuseness $a_s^*=a_d^*$ for the n+A systems was taken to be equal to that for the p+A systems. The radius parameter of the imaginary potential $r_s=r_d$ was fixed according to CH89, while and the parameter $\alpha$ was taken to equal the CH89 volume integral $J_I$ calculated with $a_s^*=a_d^*$ at 60 MeV (see Table 1).

The parameters $\beta_{I,s}$ in Eq. (7) were found here with the view of providing a description of the available data on the energy dependence of the volume integrals $J_{I,s}$ (Mahaux and Sartor, 1991b; Johnson and Mahaux, 1988; Mahaux and Sartor, 1994; Wang et al., 1993; Roberts et al., 1991; Mahaux and Sartor, 1989 and the references therein) for the n,p+$^{40}$Ca,$^{90}$Zr,$^{208}$Pb systems. The Fermi energy $E_F$ was determined to be the half-sum

$$E_F = 1/2(E_+ + E_-), \tag{9}$$

where $(-E_+)$ is the separation energy of the (A+1)-nucleon system; $(-E_-)$ is the separation energy of the A-nucleon system. The separation energies were taken from (Wapstra and Audi 1985).

The depths of the volume and surface parts of the imaginary potential, $W_s$ and $W_d$, were calculated from the known respective volume integrals $J_s^{JM}$ and $J_d^{JM}$ and from the known geometry parameters $r_s^{CH89}=r_d^{CH89}$ and $a_s^*=a_d^*$:

$$W_s(E) = \frac{J_s^{JM}(E)}{\int f(r, r_s^{CH89}, a_s^*) d\mathbf{r}}, \quad W_d(E) = \frac{J_d^{JM}(E)}{4a_d \int \frac{df(r, r_d^{CH89}, a_d^*)}{dr} d\mathbf{r}}. \tag{10}$$

The depths of the surface and volume parts of the dispersive terms were calculated in the same manner:

$$\Delta V_s(E) = \frac{J_{\Delta V_s}(E)}{\int f(r, r_s^{CH89}, a_s^*) d\mathbf{r}}, \quad \Delta V_d(E) = \frac{J_{\Delta V_d}(E)}{4a_d \int \frac{df(r, r_d^{CH89}, a_d^*)}{dr} d\mathbf{r}}. \tag{11}$$

To find the Hartree-Fock component, the optical-model code SPI-GENOA (Perey) was modified to suit the dispersive OM analysis. First, we used CH89 (with the $a_s^*=a_d^*$ values borrowed from (Romanovsky et al., 1998)) to calculate model elastic scattering differential cross sections for the

n,p+$^{40}$Ca, $^{90}$Zr, $^{208}$Pb systems in the 20-60 MeV nucleon energy range. Then, the resultant model cross sections were analyzed then in terms of the dispersive OM. In such a manner, we avoided an additional spread of the parameters that arises from the experimental errors and transformed CH89 into the dispersive potential in the 20-60 MeV nucleon energy range. The diffuseness $a_{HF}$ = 0.69 fm, the spin-orbit potential, and the Coulomb potential were fixed according to the CH89 parameterization. The imaginary potential was calculated using Eqs. (7,10), and the dispersive components using Eqs. (8,11). After that, the free parameters $V_{HF}$ and $r_{HF}$ were determined by the grid-search procedure. Namely, $V_{HF}$ and $r_{HF}$ were varied to fit the model elastic differential cross sections at nucleon energies of 20, 30, 40, 50, and 60 MeV, whereupon the averaged $r_{HF}$ was fixed and the depth $V_{HF}$ was varied again.

At $E<0$, the depth $V_{HF}$ was adjusted to reproduce the experimental single-particle energies $E_{nlj}^{exp}$ of the bound states with quantum numbers $n, l, j$ by solving the Schrodinger equation:

$$\left[\frac{-\nabla^2}{2m}+V(r,E_{nlj})\right]\Phi_{nlj}(\mathbf{r})=E_{nlj}\Phi_{nlj}(\mathbf{r}), \qquad (12)$$

where $V(r,E_{nlj})$ is the real part of dispersive OM potential (3), $\Phi_{nlj}(\mathbf{r})$ is the nucleon wave function for the orbits with quantum numbers $n, l, j$:

$$\Phi_{nlj}(\mathbf{r}) = \frac{u_{nlj}(r)}{r}Y_{lm}(\Omega). \qquad (13)$$

For the $^{40}$Ca, $^{90}$Zr, and $^{208}$Pb nuclei, we used the energies $E_{nlj}^{exp}$ determined in (Volkov et al., 1990; Vorobyev et al., 1995; Bespalova et al., 2001a) and presented in Tables 7.3 and 7.4 of (Mahaux and Sartor, 1991a) and in Tables 5 and 6 of (Mahaux and Sartor, 1991b) and calculated the energies $E_{nlj}$ by a subroutine from the DWUCK4 distorted wave code (Kunz).

Then, the depth $V_{HF}$ was parameterized for positive and negative energies by the exponential function:

$$V_{HF}(E)=V_{HF}(E_F)\exp\left(\frac{-\gamma(E-E_F)}{V_{HF}(E_F)}\right). \qquad (14)$$

Table 1 presents the resultant dispersive OM potential parameters for the n,p+$^{40}$Ca, $^{90}$Zr, and $^{208}$Pb systems.

It is well-known that, to describe the total cross sections in the low-energy range, the radius parameter $r_d$ should be made to slightly increase, and the diffuseness parameter $a_d$ to decrease with falling nucleon energy (Johnson and Mahaux, 1988). Here, the expressions of Wang et al. (1992) were used to parameterize the energy dependences of the geometrical parameters of the imaginary surface potential at low positive energies:

$$r_d(E) = r_d^{(1)} - \frac{r_d^{(2)}(E-E_F)^4}{(E-E_F)^4 + (r_d^{(1)})^4}, \qquad (15)$$

$$a_d(E) = a_d^{(1)} + \frac{a_d^{(2)}(E-E_F)^4}{(E-E_F)^4 + (a_d^{(3)})^4}. \qquad (16)$$

We found, that the expressions (15-17) with the parameters:

$r_d^{(1)} = 1.5$ fm, $r_d^{(2)} = (1.52\text{-}r_d^{CH89})$ fm, $r_d^{(3)} = a_d^{(3)} = \beta_I$=12.5 MeV,

$a_d^{(1)} = 0.1$ fm, $a_d^{(2)} = a_d^{CH89*}$- 0.05 fm (17)

are useful in the energy range where $r_d^{(15\text{-}17)} > r_d^{CH89}$ and $a_d^{(15\text{-}17)} < a_d*$ and make it possible to properly describe the available experimental data on proton reaction cross sections for the nuclei treated in low-energy range. This will be illustrated in Section 4 below. The energy dependences of $r_d$ and $a_d$ violate the sameness of the geometrical shapes of the potentials $W_d(r,E)$ and $\Delta V_d(r,E)$, between which the dispersion relation holds. Here, we neglect the violation for a simplicity.

## 3. The real potential of Woods-Saxon shape from –60 to +60 MeV

The real part of the dispersive OM potential (3) is represented by a sum of two terms with Woods-Saxon shapes (corresponding to the Hartree-Fock potential and to the volume dispersive component) and a term with a derivative of the Woods-Saxon shape (corresponding to the surface dispersive component), while the OM potential, whose real part is of the traditional Woods-Saxon shape can conveniently be used in various applications. Our aim was to obtain the global OM potential, whose real part is of Woods-Saxon form and depends on energy similarly to that of the real part of the dispersive OM potential. With that purpose, we calculated the effective Woods-Saxon real potential corresponding to the real part of dispersive OM potential from Table 1 by the technique proposed in (Mahaux and Sartor, 1989). The diffuseness of the effective real potential was taken to equal the diffuseness of the CH89 real potential $a_V^{eff} = a_V^{CH89} =$ 0.69 fm. As to the CH89 real potential depth $V^{eff}$ and radius $r_V^{eff}$, they were determined requiring that the volume integral $J_V^{eff}$ and the effective potential at $r$=0 $V^{eff}(r=0)$ should be the same as those of the real part of the dispersive OM potential . The resulting radius parameter $r_V^{eff}$ displays a characteristic wiggle in the energy range near $E_F$ (see Fig.1a).

Then, we related the calculated effective Woods-Saxon real potential to the real part of the sought potential and fixed its diffuseness to be $_V$=0.69 fm. The surface dispersive term makes the major contribution to the energy dependence of the radius parameter $r_V^{eff}$. So, we presented the radius parameter $r_V$ of the sought real potential to be a sum of the radius parameter $r_{HF}$ and the term that depends on energy in the same way as the volume integral of the surface dispersive component (8), so that:

$$r_V(E) = r_{HF} + r_V^{(1)} \left\{ \frac{\beta_I (E-E_F)\left[(E-E_F)^2 + \beta_I{}^2\right]}{(E-E_F)^4 + \beta_I{}^4} - \frac{\beta_s (E-E_F)\left[(E-E_F)^2 + \beta_s{}^2\right]}{(E-E_F)^4 + \beta_s{}^4} \right\} - \quad (18)$$
$$- r_V^{(2)} (E-E_F)^2.$$

The last term in (18) permits a better description of $r_V^{eff}(E)$ in the energy range $|E - E_F| > 40$ MeV. The parameters $r_V^{(1)}$ and $r_V^{(2)}$ were found when the chi squared, $\chi^2$, reached its minimum for the difference between $r_V^{eff}(E)$ and $r_V(E)$ (18) in each of the n,p+$^{40}$Ca,$^{90}$Zr, and $^{208}$Pb systems (see Table 2).

The depth parameter $V^{eff}$ is a monotone function of energy (see Fig. 1b), just as the Hartree-Fock component, which is described by an exponential function reasonably well. Thus, the depth parameter of the sought Woods-Saxon real potential was parameterized as

$$V = V_0 + 0.299 E_C \pm V_t \frac{N-Z}{A} + V_e \exp\left[-\kappa(E-E_F)\right] \text{ (+ for p, - for n).} \quad (19)$$

We have chosen the widely used value of the nuclear asymmetry term $V_t$ = 24 MeV. Satchler (1969) has studied the results of diverse analyses of the proton scattering from some stable nuclei at proton energies from 9 MeV to 61.4 MeV and found the mean $V_t$ value to be about 24 MeV. Varner et. al. (1991) suggested that $V_t$ = 13.1 MeV should be preferred in the case of unstable nuclei. The Coulomb correction term $0.299E_C$ ($E_C = \frac{1.73Z}{1.238A^{1/3} + 0.116}$ MeV for protons and 0 for neutrons) was taken from CH89. The least squares $\chi^2$ method was used to find the parameters $V_0$, $V_e$, and $\kappa$ by fitting $V$ (19) to $V^{eff}$ for n,p+$^{40}$Ca, $^{90}$Zr, and $^{208}$Pb systems in the energy range from –60 to 60 MeV. Table 2 presents the best fit parameters. As an example, Fig. 1 presents the dependences $r_V(E)$ and $V(E)$ (18,19) with the best fit parameters from Table 2 as compared with $r_V^{eff}(E)$ and $V^{eff}$(E) for the n+$^{90}$Zr system.

Reasonably, the parameter $r_V^{(1)}$ in Eq. (18) is expected to correlate with the parameter $\alpha$ in Eqs. (7,8) because the two parameters define the magnitude of the wiggle crest, which is characteristic of $r^{eff}(E)$ and $J_{\Delta V_I}(E)$. The parameter $r_V^{(1)}$ was found to depend almost linearly on $\alpha$. The best-fit parameters from Table 2 were averaged, resulting in that the real well depth $V$ of the nucleon-nucleus potential over the $40 \le A \le 208$ target mass and –60 to +60 MeV nucleon energy ranges has been expressed as

$$V = 25.5 + 0.299 E_C \pm 24 \frac{N-Z}{A} + 27.6 \exp\left[-0.0105 E\right] \text{ (+ for p, - for n)} \quad (20)$$

and that the parameters of the expressions (7,18) prove to be

$$\alpha = J_I^{CH89*}(60 \text{ MeV}) \; (MeV \cdot fm^3), \qquad \beta_I = 12.5 \; MeV, \qquad \beta_s = 60.0 \; MeV,$$

$$r_{HF} = 1.21 \; fm, \qquad r_V^{(1)} = 0.015 + 0.00047\alpha \;\; fm, \qquad r_V^{(2)} = 3.8 \cdot 10^{-6} \; fm/MeV^2. \tag{21}$$

In Eq. (21), the asterisk stand for the CH89 with the diffuseness $a_s$*=$a_d$* modified for a given nucleus according to (Romanovsky et al., 1998), or with the nuclei-averaged diffuseness $a_s$*=$a_d$*=0.63 fm.

The Fermi energy $E_F$ leaves but a single parameter in Eqs. (7,18) to be found separately for each given nucleus. The energy $E_F$ can be calculated using Eq. (9), or defined it in terms of centroid energies:

$$E_F = \frac{\langle E_p \rangle + \langle E_h \rangle}{2}, \quad \langle E_p \rangle = \frac{\sum_{nlj} E_{nlj}}{N_p}, \qquad \langle E_h \rangle = \frac{\sum_{nlj} E_{nlj}}{N_h}, \tag{22}$$

where the summation is with respect to the $N_p$ and $N_h$ subshells in the particle (p) and in the hole (h) valence shells. One can also use the empirical dependence of $E_F$ on relative neutron excess $N$-$Z/A$ found by Jeukenne et al. (1990) for the $40 \le A \le 208$ nuclei.

In Figs. 2 and 3, the volume integrals $J_V$ of the real potential (18,20,21) for the n,p+$^{40}$Ca, $^{90}$Zr, and $^{208}$Pb systems are compared with $J_V^{CH89}$, with the volume integrals of the global real potential found by Bauer et al. (1982), and with the results of the various individual OM analyses, including the volume integrals obtained in (Delaroche et al., 1989; Chiba et al., 1992; Wang et al., 1993; Roberts et al., 1991; Finlay et al., 1989; Perey and Perey, 1976) and the results of the present analysis of the dispersive OM. The volume integral $J_V$ of the real potential (18,20,21) is close to $J_V^{CH89}$ throughout the nucleon energy range, except for the 10-20 MeV interval, where $J_V > J_V^{CH89}$ because the CH89 does not include the dispersion relation between the real and imaginary parts of the OM potential. For medium weight nuclei, the effect of the dispersion relation on the volume integral of the real part of the nucleon-nucleus OM potential gets noticeable again in the 10-20 MeV energy range. It should be noted that the close agreement of $J_V$ with $J_V^{CH89}$ at E>20 MeV has resulted from the fact that the above mentioned model differential elastic scattering cross sections calculated using CH89 instead of experimental cross sections were analyzed in terms of the dispersive OM. Note also that the real potential (Bauer et al., 1982) was defined in the energy range less the region near the Fermi energy.

## 4. Calculation of the scattering and bound state data

In this section, we verify the predictive power of the global potential (2,10,15-18,20,21). The potential was used to calculate the scattering data for the n+$^{40}$Ca, p+$^{54}$Fe, $^{58}$Ni systems and the single-particle energies $E_{nlj}$ of proton bound states for the p,n+$^{58}$Ni,$^{116}$Sn systems. The spin-orbit potential and the imaginary potential radius were taken from CH89. The diffuseness $a_s$*= $a_d$*= 0.58, 0.60, and 0.67 fm for, respectively, $^{54}$Fe, $^{58}$Ni, and $^{116}$Sn was borrowed from (Romanovsky et al. 1998). The Fermi energy $E_F$ = ?10.6, ?8.3, ?7.0, ?5.8, and ?6.8 MeV for, respectively, the n+$^{58}$Ni, $^{116}$Sn and p+$^{54}$Fe, $^{58}$Ni, and $^{116}$Sn systems was calculated using Eq. (4).

Fig. 4 compares the calculated differential cross sections for elastic scattering of neutrons on $^{40}$Ca at 5.3, 5.9, 6.5, and 7.9 MeV with the experimental data (Reber and Brandenberger, 1967), the compound elastic contribution (Johnson and Mahaux 1988) being subtracted. Fig. 5 compares the calculated differential elastic scattering cross sections for the p+$^{58}$Ni system in the 20-60 MeV range with the experimental data from (Van Hall et al., 1977; Fricke et al., 1967; Fulmer et al., 1969; Sakaguchi et al., 1982) and with the data predicted by CH89. A good agreement has been attained at low and intermediate energies.

Fig. 6 (a,b) compares the calculated total neutron interaction cross sections $\sigma_t$ for $^{40}$Ca and total proton reaction cross sections $\sigma_r$ for $^{54}$Fe with the experimental data (Camarda et al., 1986; Carlson, 1996) and with the evaluated data (Romanovsky et al., 1995) and shows also a good agreement.

The single-particle energies of the bound states were calculated by trial-and-error method. In the procedure, some initial value $E_0$ (close to the assumed $E_{nlj}$) was chosen to use in calculating $V(E_0)$ and $r_V(E_0)$ (18,20,21) and solving the Schrodinger equation (12). After that, we calculated $\Delta_{nlj}^{(0)} = E_{nlj}^{(0)}(E_0) - E_0$, where $E_{nlj}^{(0)}(E_0)$ is the eigenvalue of the bound-state problem. Then, another energy $E_1$ was chosen to use in calculating $V(E_1)$, $r_V(E_1)$, $E_{nlj}^{(1)}(E_1)$, and $\Delta_{nlj}^{(1)}$, etc. The procedure was stopped at $\left|\Delta_{nlj}^{(n)}\right|$<10 keV after n iterations. The calculated $E_{nlj}^{(n)}(E_n)$ value was regarded as the single-particle energy of the bound state. The results for the p,n+$^{58}$Ni, $^{116}$Sn systems are listed in Tables 3 and 4 below, where they are compared with the results of joint evaluation of the stripping and pickup reaction data (Bespalova *et al* 2001b; Bespalova *et al* 2002; Boboshin 2002) and with the predictions of the relativistic mean field theory (RMFT) (Typel and Wolter 1999). A good agreement with the available experimental and theoretical data was achieved, except for the $1s_{1/2}$ state in the cases where this state is deeper than –60 MeV. Note that the applicability scope of the presented potential on the negative energy side is limited here to the range $E$ > –60 MeV.

**5. Discussion**

The presented potential (2,10,15-18,20,21) corresponds to the local equivalent energy-averaged generalized optical potential (see (Mahaux and Sartor, 1991a) for the details) and is featured mainly by a broad energy range of its applicability, including the near-Fermi energies. The true global potential is difficult to determine near the Fermi energy because the dispersive effects depend on the properties of a specific nucleus. The width $\Gamma_{nlj}$ of the quasi-particle peaks is directly related to the imaginary part of the mean field. The dependences of the empirical widths $\Gamma_{nlj}$ on the difference ($E_{nlj}$-$E_F$) for many of the $90 \le A \le 208$ nuclei are described semi-quantitatively by the unified parabola $0.04(E_{nlj} - E_F)^2$ at –15<$E_{nlj}$ -$E_F$<+15 MeV (see Fig. 7.20 in (Mahaux and Sartor 1991a)). The introduction of the global imaginary potential (and, hence, the global dispersive term) near the Fermi energy is, therefore, quite justified.

The level properties of a specific nucleus are evidently reflected by the Fermi energy, which is the parameter of the real and imaginary parts of the presented global potential. Besides, the diffuseness parameter $a_d^*=a_s^*$ , which was determined by Romanovsky et al. (1998) for a number of nuclei from $^{40}$Ar to $^{208}$Pb, may be used to allow, to a certain extent, for the properties of the specific nucleus. In their turn, the parameter $\alpha$ and the radius parameter of the real part of the potential $r_V$ depend on $a_d^*=a_s^*$.

It is known, that the OM potential is used at positive energies to calculate the energy-averaged cross sections. That is why the OM potential is little applicable to the low-energy range, where the fluctuations due to the individual states of the compound system play an important role. Actually, the OM potential (2,10,15-18,20,21) can be used in the low-energy range far from resonances.

To summarize, the main features of the presented global nucleon-nucleus OM potential are briefly as follows:

1) the Woods-Saxon real part reflects the dispersive relation via the radius parameter $r_V$ , which depends on energy similarly to the surface dispersive term of dispersive OM potential, i.e. strongly in the near-Fermi energy range;

2) the real well depth $V$ is an exponential function of energy;

3) the diffuseness of the real potential $a_V$ =0.69 fm equals the CH89 diffuseness and is constant for all the $40 \le A \le 208$ nuclei;

4) the energy dependence of the imaginary potential is described by the JM formulas with the parameter $\alpha$ calculated using CH89 parameterization (with $a_s$*= $a_d$*), $\beta_I$=12.5 MeV and $\beta_s$=60 MeV;

5) the radius parameter $r_s = r_d$ of the imaginary potential is taken from CH89; the diffuseness parameter $a_s$*= $a_d$* differs from that of CH89 and is on the average equal to 0.63 fm for the $40 \le A \le 208$ nuclei. The individual $a_s$*= $a_d$* values obtained by Romanovsky et al. (1998) for some nuclei from $^{40}$Ar to $^{208}$Pb can also be used. In the low-energy range, the radius parameter $r_d$ increases slightly, while the diffuseness $a_d$ decreases;

6) The Coulomb and spin-orbit potentials are the same as implied by CH89.

The single-particle properties of the scattering and bound-state data for the $40 \le A \le 208$ nucleon+nucleus systems at energies ranging from –60 MeV to +60 MeV can readily be evaluated using the global OM potential (2,10,15-18,20,21).

The presented potential was determined by analyzing the experimental and model data for a few nuclei, namely $^{40}$Ca, $^{90}$Zr, and $^{208}$Pb. The range of the nuclei to be studied was restricted by the scanty available experimental data on the single-particle energies of the deep bound states. An extension of the analyzed database may somewhat change the parameter values of the potential. We plan to study the uncertainties of the proposed global OM potential and verify its predictive power as regards to the nuclei with a neutron (proton) excess.

Acknowledgements. The authors wish to thank Dr. S.Typel for his providing them with the numerical results of RMFT calculations of the single particle energies of nucleon bound states in various nuclei.

Table 1

The dispersive OM potential parameters ($a_{HF}$=0.69 fm)

| System | $E_F$, MeV | $\alpha$, MeV·fm$^3$ | $\beta_I$, MeV | $\beta_s$, MeV | $r_s = r_d$, fm | $a_s$= $a_d$, fm | $V_{HF}(E_F)$ MeV | $\gamma$ | $r_{HF}$, fm |
|---|---|---|---|---|---|---|---|---|---|
| n+$^{40}$Ca | -12.0 | 103.5 | 12.0 | 60.0 | 1.207 | 0.60 | 57.05 | 0.477 | 1.207 |
| n+$^{90}$Zr | -9.88 | 81. | 12.0 | 60.0 | 1.236 | 0.61 | 53.42 | 0.428 | 1.206 |
| n+$^{208}$Pb | -5.65 | 71. | 16.0 | 40.0 | 1.259 | 0.60 | 47.16 | 0.387 | 1.230 |
| p+$^{40}$Ca | -4.71 | 103.5 | 12.5 | 60.2 | 1.207 | 0.60 | 56.32 | 0.490 | 1.207 |
| p+$^{90}$Zr | -6.73 | 100. | 12.0 | 60.0 | 1.236 | 0.61 | 59.60 | 0.473 | 1.220 |
| p+$^{208}$Pb | -5.9 | 100. | 16.0 | 70.0 | 1.259 | 0.67 | 62.39 | 0.477 | 1.230 |

Table2.

The best fit parameters of the expressions (18,19) for the n,p+$^{40}$Ca, $^{90}$Zr, and $^{208}$Pb systems

| System | $V_0$, MeV | $V_e$, MeV | $\kappa$, MeV$^{-1}$ | $r_V^{(1)}$, fm | $r_V^{(2)}$ x10$^6$, fm·MeV$^{-2}$ |
|---|---|---|---|---|---|
| n+$^{40}$Ca | 25.7 | 31.3 | 0.0102 | 0.066 | 5.5 |
| n+$^{90}$Zr | 24.8 | 31.4 | 0.0098 | 0.057 | 3.7 |
| n+$^{208}$Pb | 26.8 | 25.5 | 0.0102 | 0.047 | 2.7 |
| p+$^{40}$Ca | 26.1 | 27.8 | 0.0120 | 0.065 | 3.9 |
| p+$^{90}$Zr | 22.7 | 30.6 | 0.0105 | 0.061 | 4.2 |
| p+$^{208}$Pb | 16.2 | 35.4 | 0.0102 | 0.059 | 2.5 |

Table 3

Single-particle energies of the bound states ( in MeV) for the systems p,n+$^{58}$Ni

| System | Subshell | $E_{nlj}$ | $E_{nlj}^{exp}$ * | $E_{nlj}^{RMFT}$ |
|---|---|---|---|---|
| p+$^{58}$Ni | $1s_{1/2}$ | -55.7 | | -53.80 |
| | $1p_{3/2}$ | -38.4 | | -39.64 |
| | $1p_{1/2}$ | -34.2 | | -36.96 |
| | $1d_{5/2}$ | -19.6 | | -24.26 |
| | $2s_{1/2}$ | -14.3 | | -15.23 |
| | $1d_{3/2}$ | -13.8 | -11.7(8) | -18.18 |
| | $1f_{7/2}$ | -7.6 | -7.68(46) | -9.18 |
| | $2p_{3/2}$ | -3.4 | -2.46(14) | |
| | $2p_{1/2}$ | -1.9 | -0.67(7) | |
| | $1f_{5/2}$ | -1.6 | -1.15(7) | |
| n+$^{58}$Ni | $1s_{1/2}$ | -69.8 | | -63.16 |
| | $1p_{3/2}$ | -49.6 | | -48.64 |
| | $1p_{1/2}$ | -45.3 | | -46.05 |
| | $1d_{5/2}$ | -29.8 | | -32.88 |
| | $2s_{1/2}$ | -23.0 | | -23.76 |
| | $1d_{3/2}$ | -22.3 | -20.0(20) | -26.90 |
| | $1f_{7/2}$ | -15.0 | -15.3(11) | -17.29 |
| | $2p_{3/2}$ | -11.1 | -9.9(9) | -8.89 |
| | $2p_{1/2}$ | -9.43 | -9.4(8) | -7.27 |
| | $1f_{5/2}$ | -9.04 | -10.7(9) | -8.36 |
| | $1g_{9/2}$ | -4.44 | -5.8(7) | -2.88 |
| | $2d_{5/2}$ | -1.70 | -2.7(7) | |

* $E_{nlj}^{exp}$ (Bespalova et al., 2001b; Bespalova et al., 2002) were obtained by joint evaluating the stripping and pickup reaction data. The values in parenthesis are the errors due to the uncertainty of the final-nucleus state spins; a 10% experimental error should be added.

Table 4

Single-particle energies of the bound states ( in MeV) for the systems p, n+$^{116}$Sn

| System | Subshell | $E_{nlj}$ | $E_{nlj}^{exp}$ * | $E_{nlj}^{RMFT}$ |
|---|---|---|---|---|
| p+$^{116}$Sn | $1s_{1/2}$ | -62.9 | | -54.21 |
| | $1p_{3/2}$ | -50.2 | | -44.98 |
| | $1p_{1/2}$ | -48.2 | | -43.69 |
| | $1d_{5/2}$ | -36.2 | | -34.15 |
| | $2s_{1/2}$ | -29.1 | | -26.89 |
| | $1d_{3/2}$ | -32.0 | | -31.01 |
| | $1f_{7/2}$ | -19.7 | | -22.54 |
| | $2p_{3/2}$ | -14.1 | | -13.69 |
| | $1f_{5/2}$ | -15.2 | | -17.12 |
| | $2p_{1/2}$ | -12.9 | -9.8(3) | -12.18 |
| | $1g_{9/2}$ | -10.2 | -9.4(3) | -10.71 |
| | $1g_{7/2}$ | -4.8 | -4.2(5) | |
| | $2d_{5/2}$ | -4.8 | -3.7(2) | |
| | $3s_{1/2}$ | -2.8 | -3.2(1) | |
| | $2d_{3/2}$ | -2.9 | | |
| | $1h_{11/2}$ | -1.6 | | |
| n+$^{116}$Sn | $1s_{1/2}$ | -75.4 | | -64.40 |
| | $1p_{3/2}$ | -60.1 | | -54.52 |
| | $1p_{1/2}$ | -58.1 | | -53.35 |
| | $1d_{5/2}$ | -45.3 | | -43.11 |
| | $1d_{3/2}$ | -41.2 | | -40.30 |
| | $2s_{1/2}$ | -39.2 | | -36.59 |
| | $1f_{7/2}$ | -29.9 | | -31.02 |
| | $1f_{5/2}$ | -23.1 | | -25.83 |
| | $2p_{3/2}$ | -22.1 | | -22.52 |
| | $2p_{1/2}$ | -20.2 | | -21.05 |
| | $1g_{9/2}$ | -16.1 | | -18.72 |
| | $2d_{5/2}$ | -11.4 | -10.0(9) | -9.94 |
| | $1g_{7/2}$ | -10.6 | -9.9(8) | -11.26 |
| | $3s_{1/2}$ | -9.5 | -8.3(8) | -7.56 |
| | $2d_{3/2}$ | -9.3 | -7.7(8) | -7.67 |
| | $1h_{11/2}$ | -7.2 | -7.3(7) | -6.72 |
| | $2f_{7/2}$ | -2.8 | | |
| | $1h_{9/2}$ | -0.7 | | |

* The $E_{nlj}^{exp}$ values (Boboshin 2002) were obtained by joint evaluating the stripping and pickup reaction data. The values in parenthesis are the same as in Table 3.

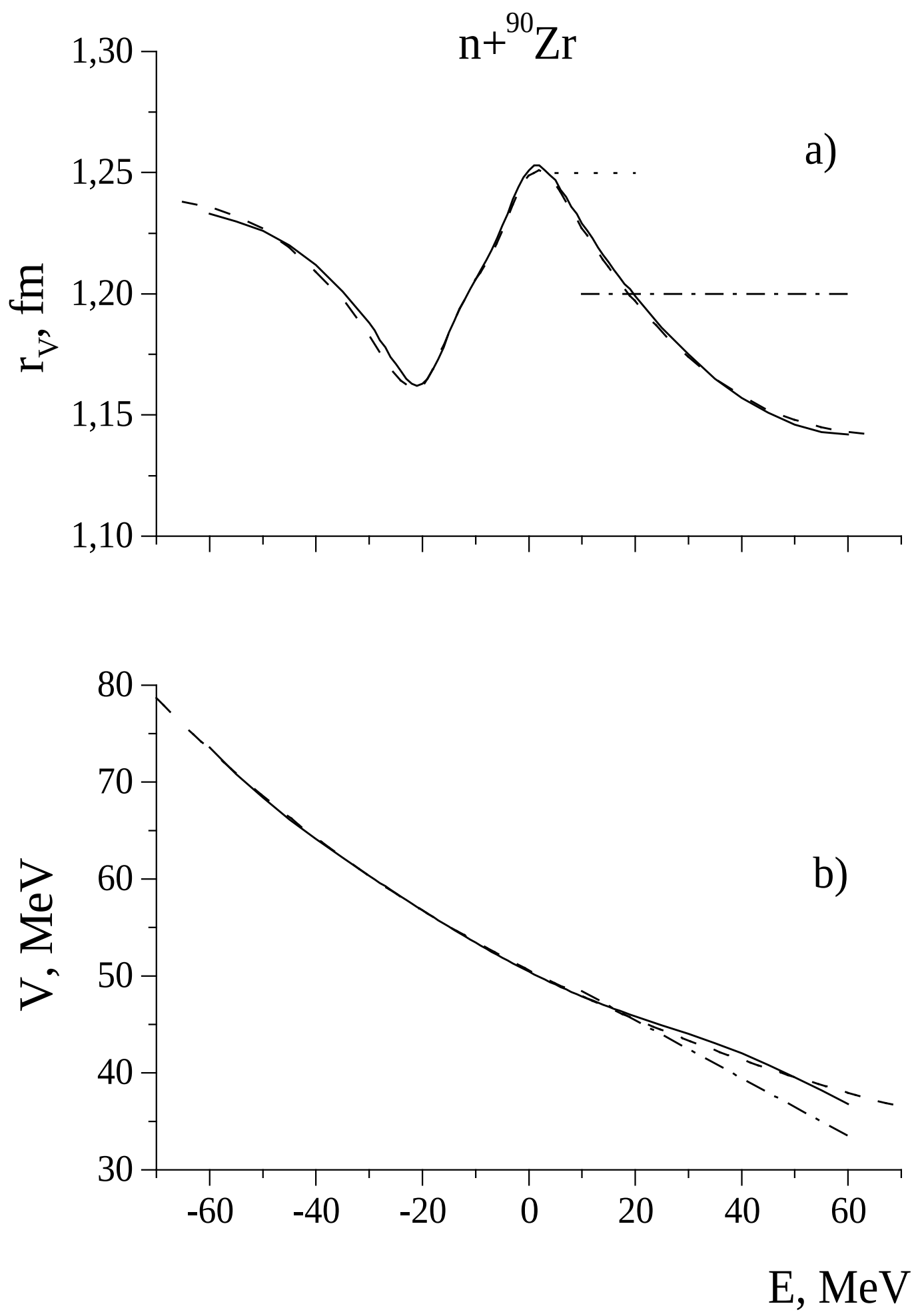

Fig. 1. The radius and depth parameters of the real potential for the n+$^{90}$Zr system.

*Panel a.* The full curve is the radius parameter, $r_V^{eff}$, of the effective real potential; the broken line is the radius parameter $r_V$ calculated using (18) with the best fit parameters from Table 2; the chain curve is $r_V^{CH89}$; the dotted curve is $r_V$ from the systematics of (Perey and Perey, 1976).

*Panel b.* The full curve is the depth parameter, $V^{eff}$, of the effective real potential; the broken curve is the depth parameter $V$ calculated using (19) with the best fit parameters from Table 2; the chain curve is $V^{CH89}$.

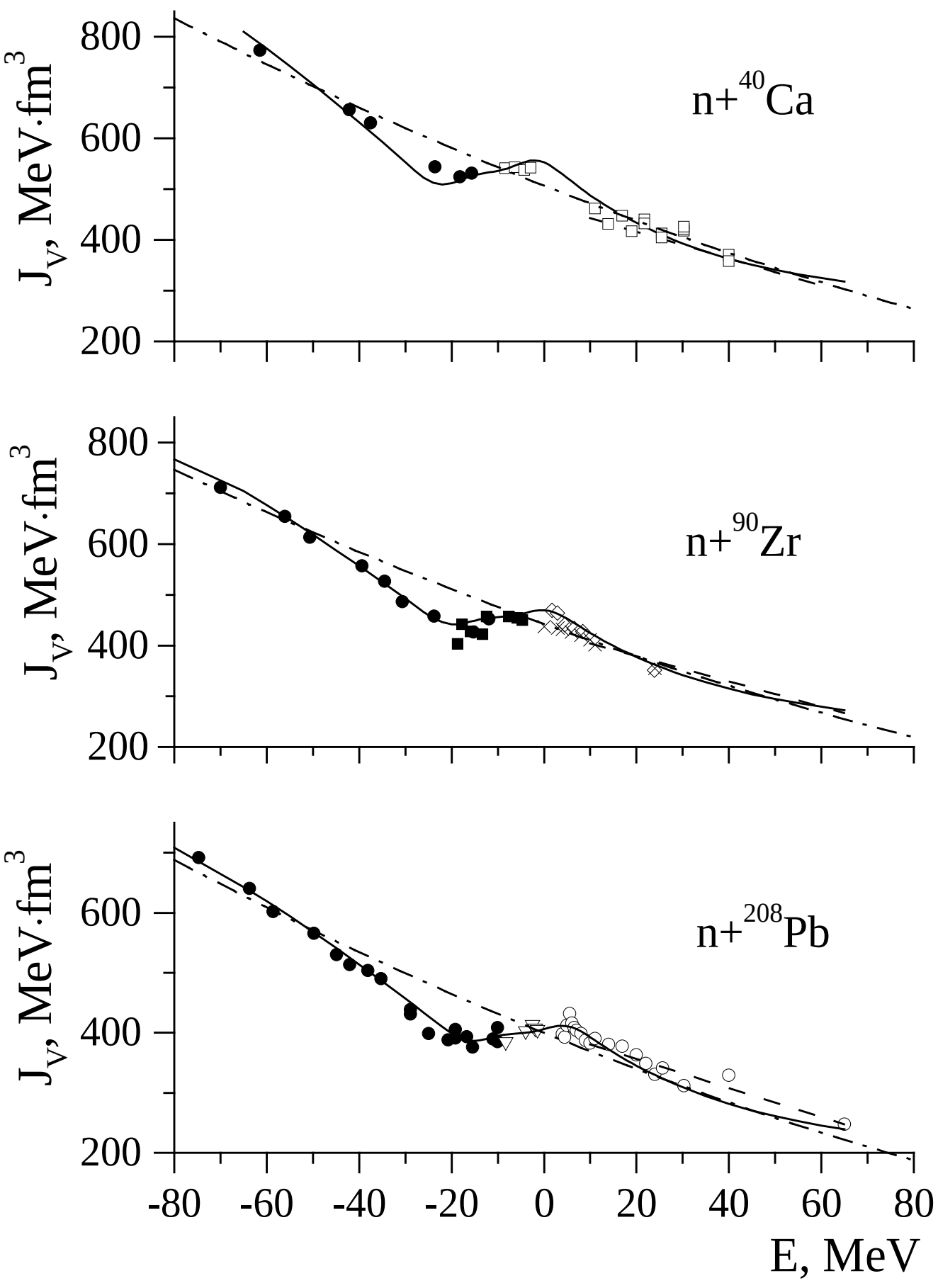

Fig. 2. The volume integrals of the real potential $J_V$ for the n+$^{40}$Ca, $^{90}$Zr, and $^{208}$Pb systems. The full curve is $J_V$ calculated using (18,20,21), the broken curve is $J_V^{CH89}$; the chain curve is $J_V$ (Bauer et al., 1982). The data points at energy $E<0$ refer to the results of our dispersive OM analysis of the energies $E_{nlj}^{exp}$ measured in (Volkov et al., 1990; Vorobyev et al., 1995) (the black circles), $E_{nlj}^{exp}$ given in Tables 7.3 and 7.4 of (Mahaux and Sartor, 1991a) (the triangles open downwards) and in Tables 5 and 6 of (Mahaux and Sartor, 1991b) (the light squares), $E_{nlj}^{exp}$ obtained by joint evaluation of the stripping and pickup reaction data (Bespalova 2001) (the black squares). The data points at $E>0$ refer to the potential given in Table 1 of (Mahaux and Sartor, 1991b) (the light squares), to the potentials of (Chiba et al., 1992) (the light rhombi), of (Delaroche *et al* 1989) (the crosses), and of (Roberts *et al* 1991) (the light circles).

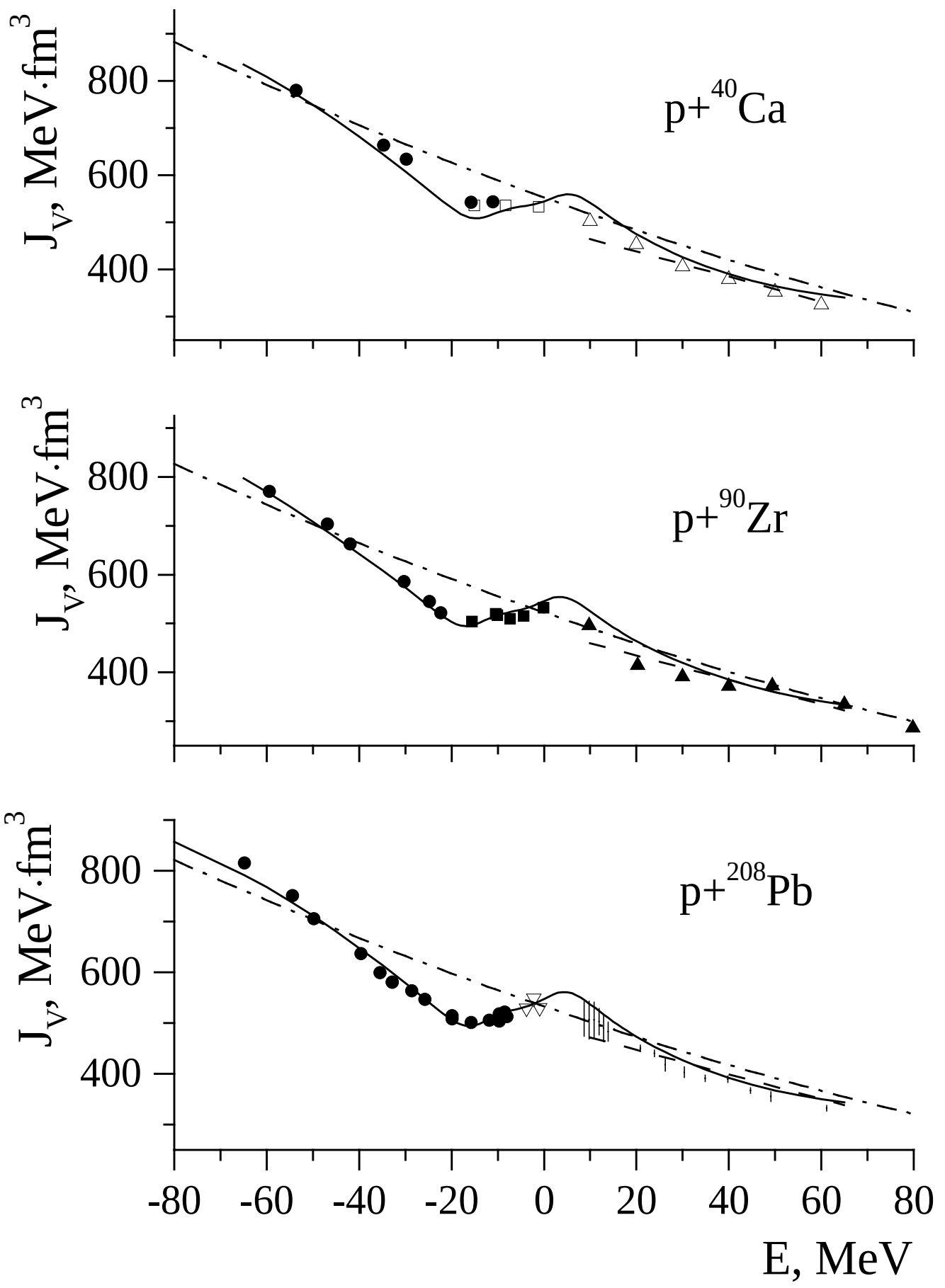

Fig.3. The same as in Fig.2 for the p+$^{40}$Ca, $^{90}$Zr, and $^{208}$Pb systems. The data dots at the energy *E>0* refer to the OM potentials contained in the compilation of Perey and Perey (1976) (the triangles open upwards), to the potential given by Wang et al. (1993) in Table I (the black triangles), and the grid-search results of Finlay et al. (1989) (the bars) .

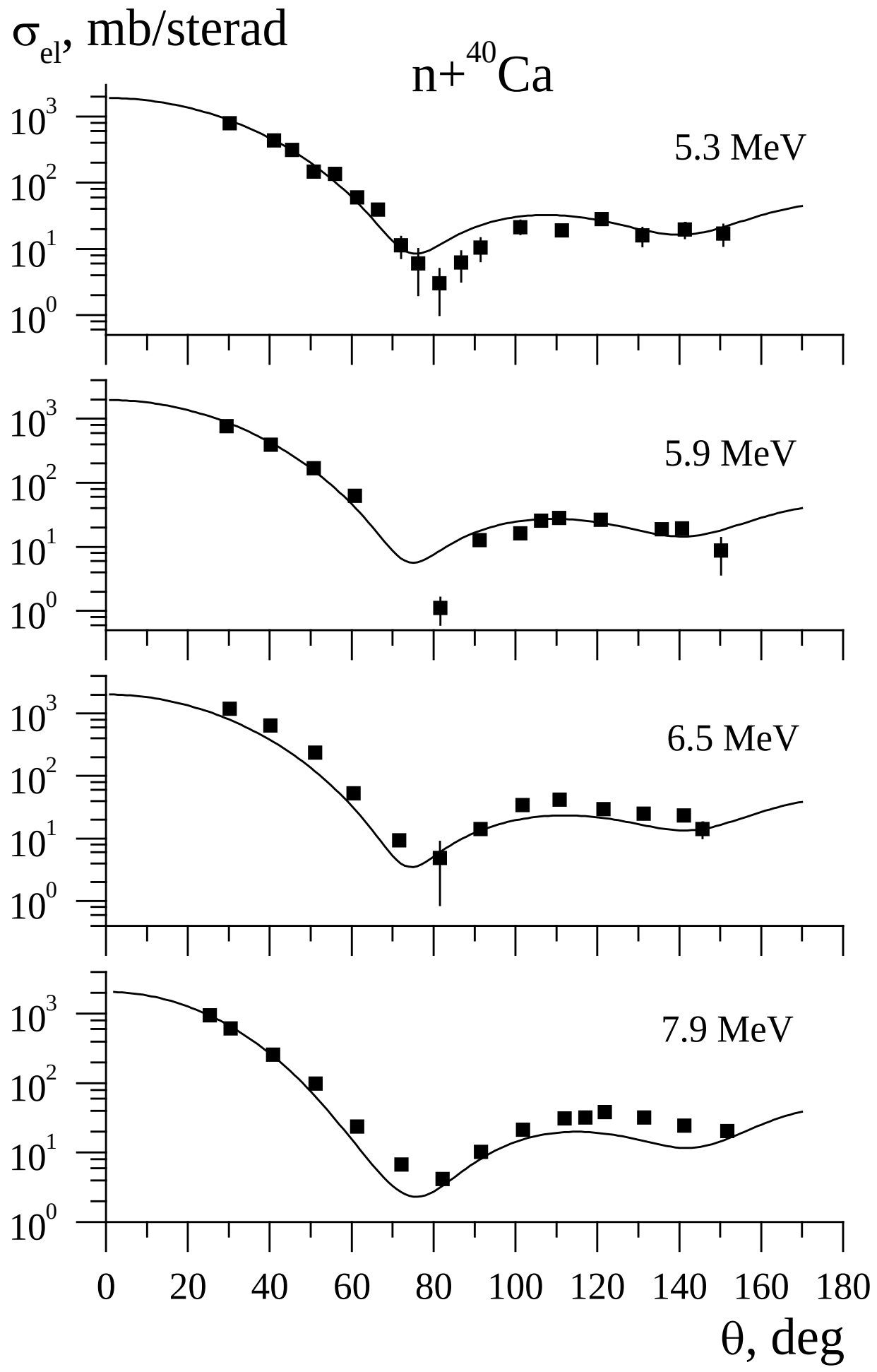

Fig.4. Differential elastic scattering cross sections for the n+$^{40}$Ca system. The black curves are the calculations with the potential (2,10,15-18,20,21). The black squares are the experimental data (Reber and Brandenberger 1967), from which the compound elastic contribution (Johnson and Mahaux 1988) have been subtracted.

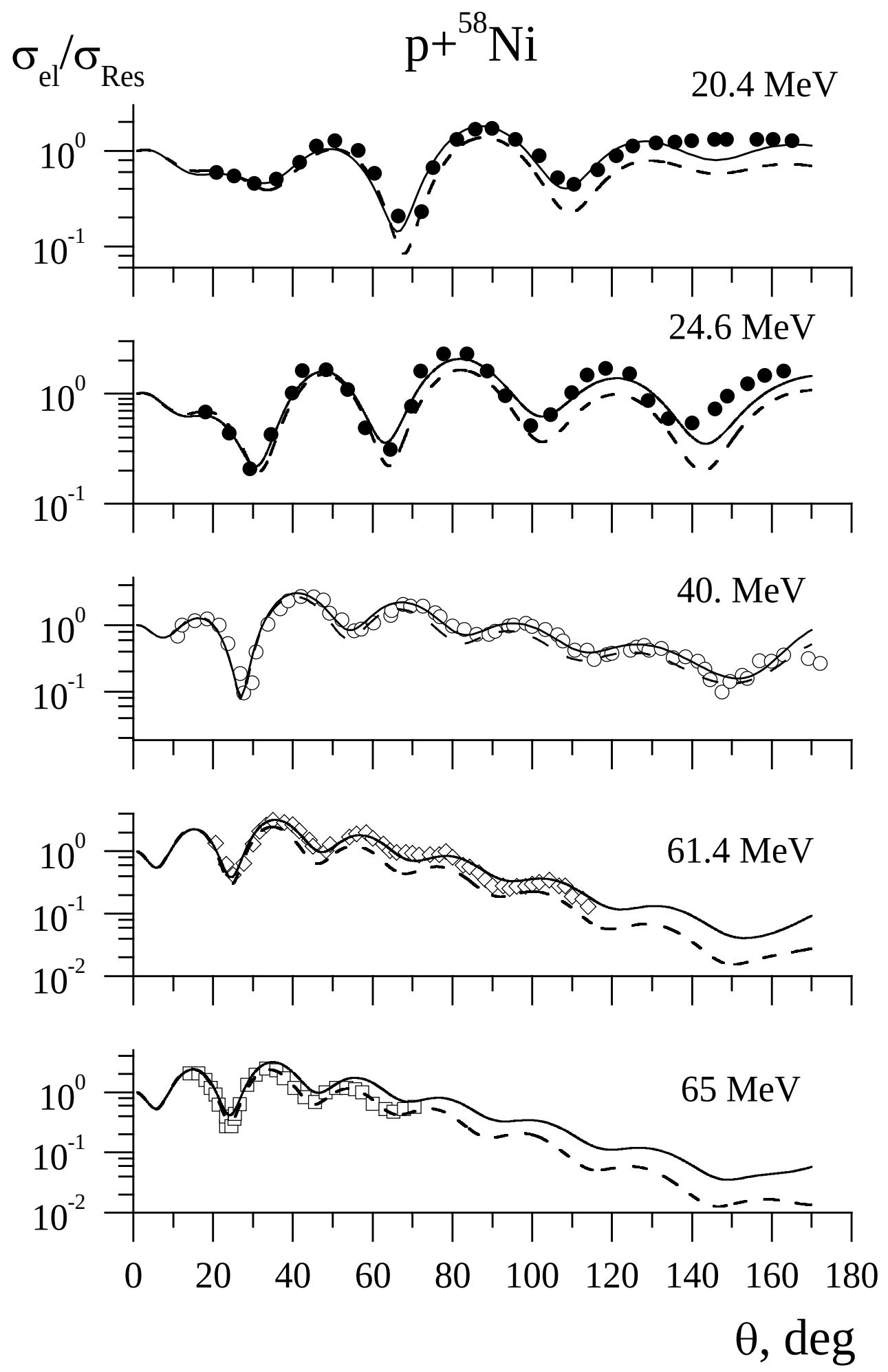

Fig. 5. Differential elastic scattering cross sections for the p+$^{58}$Ni system. The solid curves are the calculation with the potential (2,10,15-18,20,21). The broken curve shows the cross sections predicted by CH89. The markers are the experimental data of Van Hall *et al* (1977) (the black circles), of Fricke et al. (1967) (the light circles), of Fulmer et al. (1969) (the black squares), and of Sakaguchi et al. (1982) (the light squares).

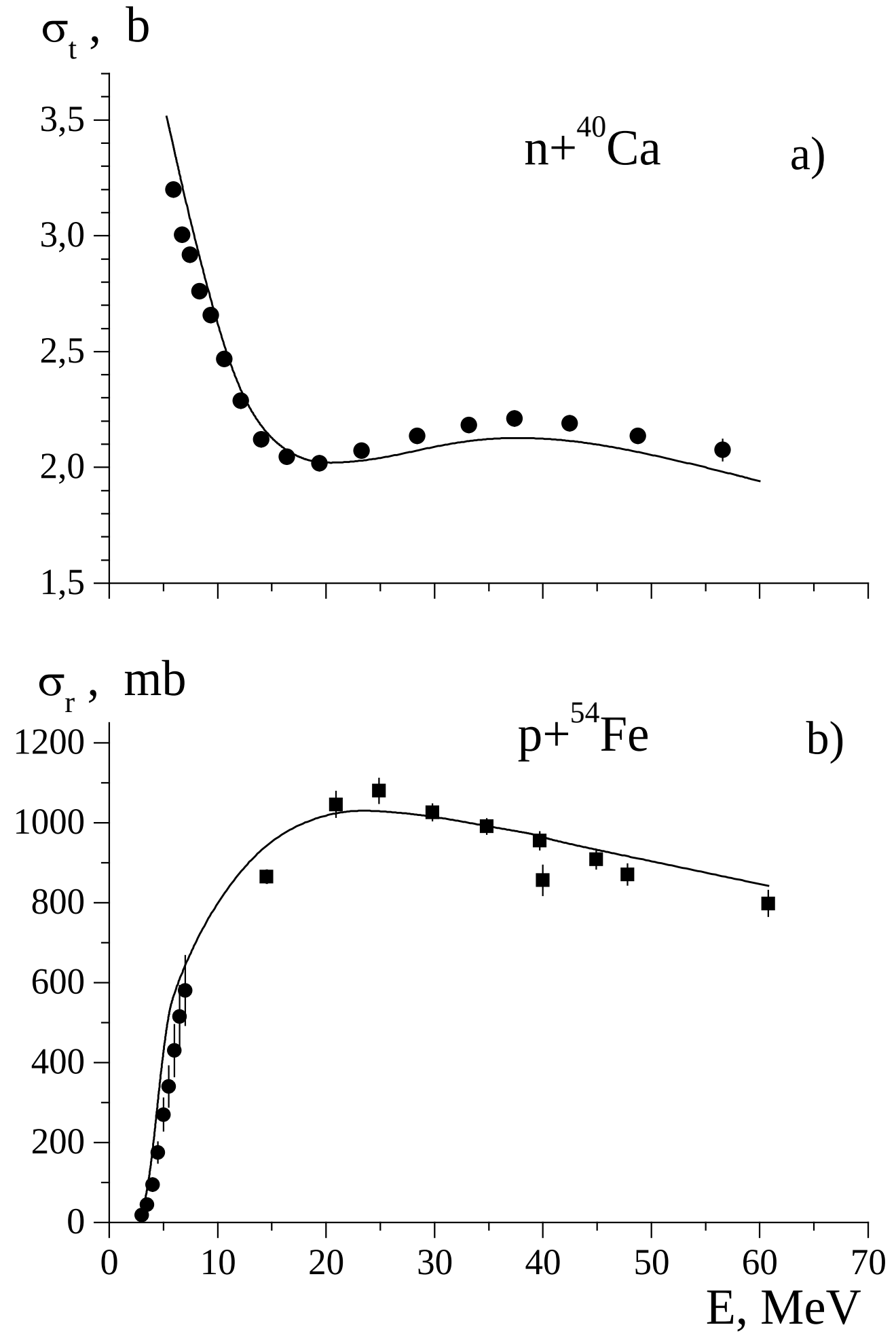

Fig.6. (a) The total neutron interaction cross sections for the n+$^{40}$Ca system. The solid curve shows the calculations with the potential (2,10,15-18,20,21). The circles correspond to the experimental data of Camarda et al. (1986). (b) The total proton reaction cross sections for the p+$^{54}$Fe system. The solid curve shows the calculation with the potential (2,10,15-18,20,21). The circles are the evaluated data (Romanovsky *et al* 1995). The squares are the experimental data of Carlson (1996).